\documentclass{PoS}

\title{Non-perturbative QCD effects in $\eta_c$ and 
$\eta_b$ decays into baryons and WIMP scattering off nuclei}

\ShortTitle{Non-perturbative QCD effects in $\eta_{c,b}$ decays into baryons and WIMP scattering off
nuclei}

\author{\speaker{Miguel-Angel Sanchis-Lozano}%
        \\
        IFIC-Departamento de Fisica Teorica, CSIC-Valencia University, Spain\\
        E-mail: \email{Miguel.Angel.Sanchis@uv.es}}

\author{Nikolai Kochelev\\
        BLTP, JINR, Dubna, Russia and Department of Physics,  Kyungpook National University, Daegu, 
South Korea\\}

\author{Redamy P\'erez-Ramos\\
Department of Physics, University of Jyv\"{a}skyl\"{a},  
Finland \\ }

\abstract{In this work we estimate  
the helicity suppressed
decay rates of $\eta_b$ resonances into baryon pairs
due to instanton-induced effects by rescaling
the corresponding partial widths of the experimentally measured
branching ratios for the $\eta_c(1S) \to p\bar{p}$ 
and $\eta_c(1S) \to \Lambda\bar{\Lambda}$ decay modes. Thus 
we point out that both $\eta_b(1S) \to p\bar{p}$ 
and $\eta_b(1S) \to \Lambda\bar{\Lambda}$ channels could be
detected at a Super B factory and LHC experiments. 
Furthermore, we examine related instanton-induced 
effects on WIMP scattering off nuclei concluding, albeit with large
uncertainties, that they might enhance the spin-dependent
cross section for a light pseudoscalar Higgs mediator, thereby
inducing a dependence on the momentum transfer
to the recoiling nucleus.}

\FullConference{Xth Quark Confinement and the Hadron Spectrum\\
        8~V12 October 2012\\
        TUM Campus Garching, Munich, Germany}

\begin{document}

\section{Introduction}

As is well-known, non-abelian gauge theories show non-trivial
properties associated to a tunneling process between
different classical field configurations. In particular
instanton effects can be related to the
$U(1)$-problem, $\theta$-problem and chiral symmetry breaking
in QCD, leading to many phenomenological implications in  hadron physics. 
For general reviews see, e.g., \cite{Kochelev:2005xn}, \cite{Schafer:1996wv}, \cite{Diakonov:2002fq}.

In this work
we first examine the decays of $\eta_c$  pseudoscalar resonances
(later extending our analysis to $\eta_b$ states)
into baryon pairs which pose a long-standing puzzle to perturbative
QCD, as commented below.

Let us get started by remarking that instantons can be
interpreted in either way:
\begin{itemize}

\item[{\it a)}] Instantons are large fluctuations of the gluon field
corresponding to a tunneling process occurring in time whose
amplitude is proportional to $\exp(-2\pi/\alpha_s)$, where
$\alpha_s$ denotes the strong coupling constant. It is
apparent that such effects are non perturbative and
instantons are missed in all orders of perturbation
theory.

\item[{\it b)}] Instantons are localized pseudoparticles in Euclidean
space-time that can induce interactions between quarks and
gluons.

\end{itemize}

The physical QCD vacuum should resemble a kind of 
$\lq\lq$liquid'', provided that the typical instanton size
$\bar{\rho} \simeq 1/600\ {\mathrm MeV}^{-1} = 1/3$ fm 
is smaller than the instanton separation $R \approx 1$ fm.

\subsection{Motivating ideas}

It has been found experimentally that the decay rate
of the $\eta_c$ resonance into baryon pairs is
much larger than expected according to perturbative QCD,
where such processes should be largely suppressed by
helicity conservation.

In particular, an explanation of the large observed  
decay rates of $\eta_c(1S)$ resonance to baryon pairs \cite{pdg}:
\begin{itemize} 
\item $BF[\eta_c(1S) \to p\bar{p}] = (1.43 \pm 0.17) \times 10^{-3}$
\item $BF[\eta_c(1S) \to \Lambda\bar{\Lambda}] = (0.94 \pm 0.32) \times 10^{-3}$
\end{itemize}
seems to require a fundamental modification of the
perturbative approach.

Different proposals have been put forward in terms of 
a non-leading or non-perturbative mechanism: mixing of the 
resonance and gluonium states \cite{hep-ph/9310344}, instanton effects 
\cite{Anselmino:1993bd}, 
intermediate meson loop contribution~\cite{Liu:2010um}
or higher Fock components of the hadronic resonance \cite{Feldmann:2000hs}. 
Despite many uncertainties, it is conceivable that 
long-distance contributions may also affect
$\eta_{b}(nS)$ resonances. In Fig.~1 we show the
possible instanton interaction via a couple
of gluons perturbatively coupled to the $\eta_{c,b}$ meson (left)
and non-perturbatively coupled to a baryon pair (right)
via instanton interactions. 

On the other hand, one can wonder whether a non-perturbative
contribution may have an influence in the scattering
of dark matter (e.g. WIMP) off nuclei, when the mediator 
couples to nucleons via a triangle-loop diagram with two gluons
as depicted in Fig.2.

\begin{figure}[ht!]
\begin{center}
\includegraphics*[width=.33\linewidth]{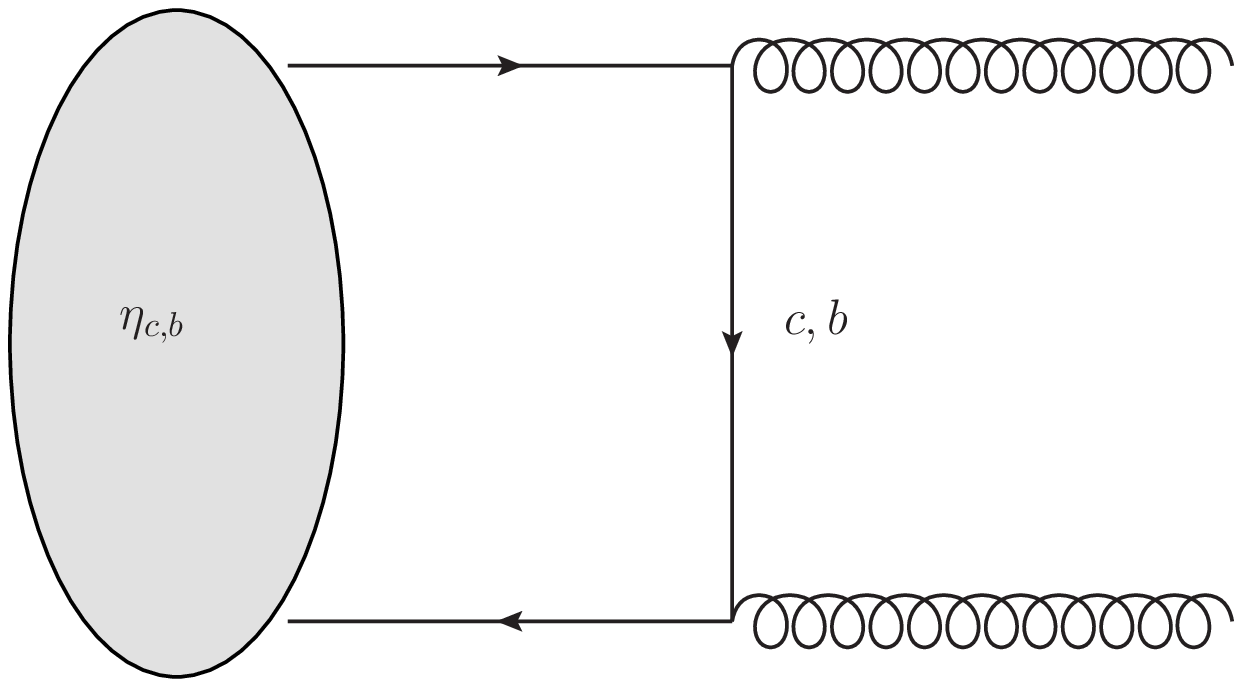}
\hskip 1.1cm
\includegraphics*[width=.4\linewidth]{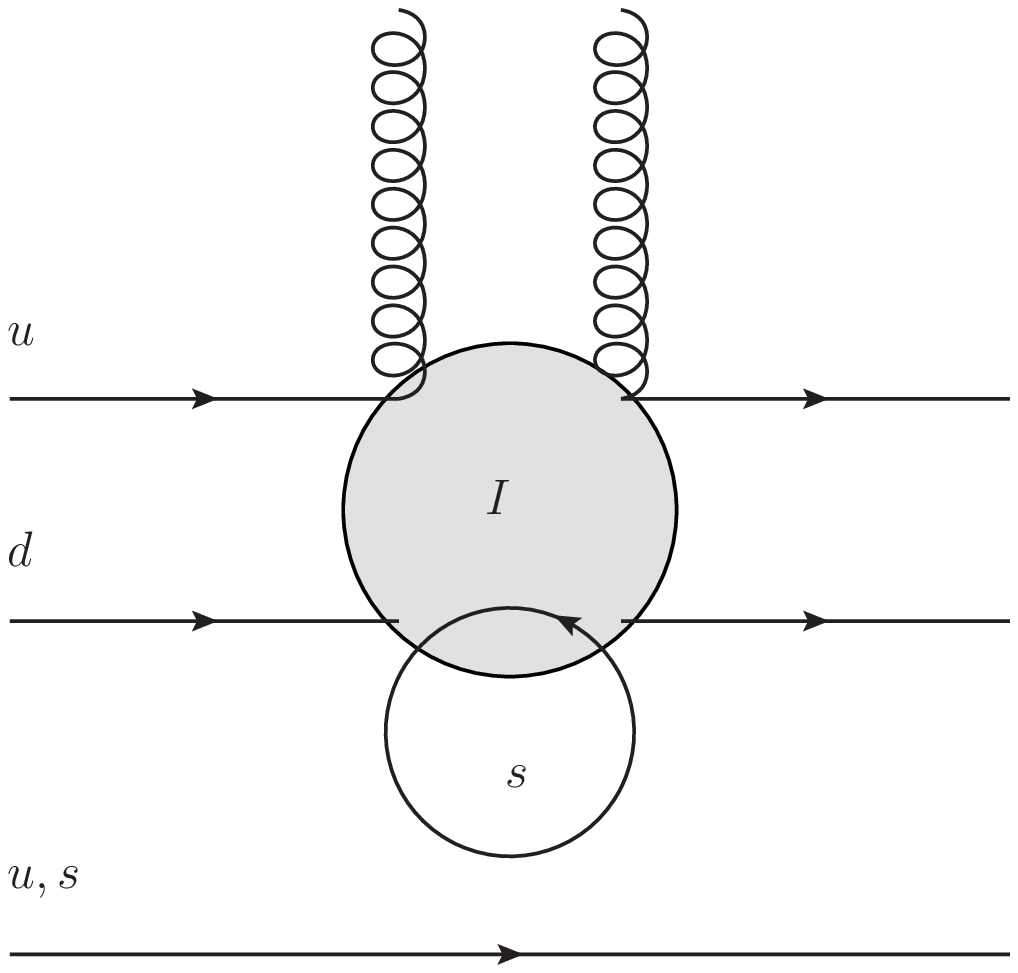}
\end{center}
\caption{Left: Perturbative coupling of
two gluons to a pseudoscalar resonance $\eta_{c,b}$. Right:
Non-perturbative coupling of a couple of gluons to a $p\bar{p}$ or to $\Lambda\bar{\Lambda}$ pairs
through an instanton-induced interaction.}
\label{fig:1}
\end{figure}

\section{Instanton size distribution}

The analogy with a a liquid pointed out
in the Introduction allows one to introduce
an instanton density with a size distribution
which can be calculated from QCD under some approximations
and regularization scheme.

At the one-loop approximation the instanton size
distribution is given by \cite{'tHooft:1976fv}: 
\begin{equation}\label{eq:exp} 
n(\rho) \sim \frac{1}{\rho^5}\ [\alpha_s(\rho^{-1})]^{-6}\
\exp{\biggl[-\frac{2\pi}{\alpha_s(\rho^{-1})}\biggr]}
\end{equation}
where
\begin{equation}\label{eq:alpha}
\alpha_s(\rho^{-1})=\frac{g_s^2(\rho^{-1})}{4\pi} \simeq 
\frac{2\pi}{b\log{(1/\Lambda\rho)}}
\end{equation}
with $b=11-2N_f/3$, where  $N_f$ denotes the
the number of active flavors;
$\Lambda$ is of the order of 250 MeV. Let us remark that $\alpha_s(\mu)$
in the pre-exponent factor of Eq.(\ref{eq:exp}) starts to run
only at the two-loop approximation; hence its argument
is set at the ultraviolet cutoff $\mu$.

Note that at small $\rho$ and $N_f=3$, the size distribution is given
approximately by:
\begin{equation}\label{eq:scaling}
n(\rho) \sim \rho^{b-5} = \rho^4
\end{equation}
i.e. it increases with increasing instanton size. Thus
the integral of $n(\rho)$ over $\rho$ diverges, representing an infrared problem.
Actually, such infrared divergence can be seen as an artifact of
using the one-loop formula for $\alpha_s(1/\rho)$ in Eq.(\ref{eq:alpha}).
Moreover, if the semiclassical approximation is meaningful at all, 
a solution to this problem has to be found in the context
of the full instanton ensemble.The simplest way is to cut
the integrations off at some given value $\rho_0$.
A dynamical cut-off could originate from configurations where
instantons start to overlap, undergoing
a repulsive interaction. 

On the other hand, there are arguments in favor of a suppression of the
type $n(\rho) \sim \exp{(-c\rho^2)}$, which is physically equivalent
to set a cutoff $\rho_0$ for large instanton sizes. We will turn to this
point in section 4.
 
\newpage

\subsection{The case of heavy quarkonium decays into baryons}

In the following we make the hypothesis that heavy quarkonium
(charmonium and bottomonium) mainly couples to instantons of size 
$\rho_{c,b} \approx (m_{c,b}v_{cb})^{-1}$, where $v_{c,b}$
denote the heavy quark velocities, respectively.
In particular we consider $\eta_{c,b}$ states with the simplifying
assumption
\begin{equation}\label{eq:delta}
n(\rho)\ =\ n_{c,b}\delta(\rho-\rho_{c,b})
\end{equation}
where $n_{c,b}$ is assumed to follow the same scaling
as Eq.(\ref{eq:scaling}). (As we shall see, heavy quarks 
in the quarkonium system 
do not play a role in the instanton-gluon interaction, although they couple
perturbatively to the gluon pair for pseudoscalar states.). Therefore
\begin{equation}
r_c\ =\ \frac{n_b}{n_c} \simeq \biggl[\frac{\rho_b}{\rho_c}\biggr]^4 \simeq 0.1
\end{equation}

\subsection{Effective instanton induced gluon-quark vertex}

The relevant Lagrangian can be written as
\begin{equation}\label{eq:lagrangian}
{\cal L}\ =\ \int d\rho\ n(\rho)\biggl[\mathrm{t'Hooft}(\gamma_5)+
\mathrm{two-gluon\ interaction}
\biggl]
\end{equation}
with the following rules in our case:
\begin{itemize}
\item[i)] Each quark in zero mode in instanton field gives a factor $\rho^3$ 
\item[ii)] Each closed quark loop gives a factor $m_{current}\rho$
\item[iii)] Each gluon coupling to the instanton field gives a factor $\rho^2/g_s(\rho^{-1})$
\end{itemize}

According to the previous rules, the lowest order diagram contributing
to the $\eta_b$ decay into a baryon pair is shown in Fig.1.
A caveat is in order however: Although diagrams with extra light 
quark loops are suppressed because of small masses of $up$ and $down$ quarks, 
such diagrams would contribute with lower $\rho$ powers; hence
the suppression might be somewhat balanced at the end. Nonetheless, 
in the following we focus on diagram of Fig.1 as the leading contribution
to the effective coupling of gluons to baryons.

The amplitude of the diagram depicted in Fig.~1 (right) in the case of the 
pseudoscalar quantum number for two gluon state
should scale with the instanton size $\rho$ according to  
(see also \cite{Zetocha:2002}):
\begin{equation}\label{tHooft} 
(\rho^3)^2 \times (m_s\rho) \times 
(\bar{u}u\bar{d}\gamma_5d+\bar{d}d\bar{u}\gamma_5u) \times 
\biggl[\frac{\rho^2}{g_s(\rho^{-1})}\biggr]^2\ G_{\mu\nu}\tilde{G}_{\mu\nu}
\end{equation}
which represents that the interaction amplitude should be 
$\sim \rho^{11}/\alpha_s(\rho^{-1})$,
to be convoluted with the instanton density $n(\rho)$, leading to
\begin{equation}\label{eq:amplitude}
\int d\rho\ n(\rho)\ 
\rho^{11}/\alpha_s(\rho^{-1})\ \sim\ \rho_{c,b}^{15}/\alpha_s(\rho_{c,b}^{-1})
\end{equation} 
in the approximation given in Eq.(\ref{eq:delta}).

\section{Numerical estimates for $\eta_b \to p\bar{p}$ and $\eta_c \to \Lambda\bar{\Lambda}$
partial widths}

We shall rescale the decay rate of the $\eta_c(1S)$ into a baryon pair $B\bar{B}$
taking into account both kinematical and dynamical factors
according to the ansatz:

\begin{equation}
\frac{\Gamma[\eta_b \to B\bar{B}]}{\Gamma[\eta_c \to B\bar{B}]}\ \sim\ 
\frac{K[\eta_b]}{K[\eta_c]} \times r_c \times r_{\rho}
\end{equation}
where
\begin{equation}
\frac{K[\eta_b]}{K[\eta_c]} = \frac{m_{\eta_b}^4}{m_{\eta_c}^4}\times 
\frac{\sqrt{1-4m_p^2/m_{\eta_b}^2}}{\sqrt{1-4m_p^2/m_{\eta_c}^2}}\times 
\biggl[\frac{R_{\eta_b(1S)}(0)}{R_{\eta_c(1S)}(0)}\biggr]^2 \sim 10^3
\end{equation}
\begin{equation}
r_c=\biggl[\frac{n_b}{n_c}\biggr]^2 \simeq \biggl[\frac{\rho_b}{\rho_c}\biggr]^8 \sim 10^{-2}\ ;\ \
r_{\rho}=\biggl[\frac{\alpha_s(m_b)}{\alpha_s(m_c)}\biggr]^2 \times
\biggl[\frac{\alpha_s(\rho_c^{-1})}{\alpha_s(\rho_b^{-1})}\biggr]^2 \times
\biggl[\frac{\rho_b}{\rho_c}\biggr]^{22}\sim 10^{-6}-10^{-5}
\end{equation}

Setting numerical values: $|{R_{\eta_b(1S)}(0)}/{R_{\eta_c(1S)}(0)}|^2 \simeq 
6.477/0.81 \simeq 8$ \cite{Eichten:1995ch}, 
$BF[\eta_c \to p\bar{p}] = (1.41 \pm 0.17)\times 10^{-3}$, 
and  $BF[\eta_c \to \Lambda\bar{\Lambda}] = (9.4 \pm 3.2)\times 10^{-4}$
\cite{pdg}, we are led to:
\begin{equation}
\frac{\Gamma[\eta_b \to B\bar{B}]}{\Gamma[\eta_c \to B\bar{B}]}\ \sim\
10^{-5}-10^{-4}\ \to\ \Gamma[\eta_b(1S) \to B\bar{B}]\ \approx\ 
{\cal O}(1-10)\ 
{\mathrm eV}
\end{equation} 
Next, making use of the total width
of the $\eta_b(1S)$ resonance, found to be 
$\Gamma[\eta_b(1S)] \simeq 10\ \mathrm{MeV}$
by Belle \cite{Collaboration:2011chb},
we are able to provide the following order-of-magnitude predictions:
\begin{equation}\label{eq:prediction}
BF[\eta_b(1S) \to p\bar{p}]\ \simeq\ 
BF[\eta_b(1S) \to \Lambda\bar{\Lambda}] \sim 10^{-7}-10^{-6} 
\end{equation}

Such values of the branching fractions
imply that $\eta_b(1S)$ decays to baryon pairs
should be within reach of LHC experiments \cite{2012} 
and a Super B Factory.

\section{Extrapolation to small transfer momentum in WIMP 
scattering off nuclei}

Direct detection of dark matter through WIMP scattering
off nuclei has become one of the hottest
(and controversial) points in physics today. Indeed, 
some experiments claim to have already found evidence 
of it whereas
others fail to detect any signal at all.
On the one hand, DAMA/LIBRA, CoGeNT, and
more recently CRESST experiments have
reported the observation of events in excess of the expected
background, hinting at the existence of a light
WIMP~\cite{Bernabei:2010mq,Aalseth:2011wp,Angloher:2011uu}.
On the other hand, exclusion limits
set by other direct searches, such as Xenon10~\cite{arXiv:1104.3088}  
and Xenon100~\cite{arXiv:1104.2549}, 
are in tension with the above claims.

The total WIMP-nucleus cross section has
contributions from both spin-independent and 
spin-dependent interactions, though
one contribution is expected to 
dominate the other depending on the
target nucleus (e.g. according to the even/odd  
number of protons and neutrons) and 
the detection technique employed in the experiment.
The contributions to the spin-independent cross section arise 
in the interaction Lagrangian of the WIMP
with quarks and gluons of the nucleon from
scalar and vector couplings whereas the spin-dependent part
is attributed to the axial-vector couplings.
Pseudoscalar interaction is usually neglected because
of a strong velocity and/or momentum transfer suppression.

\begin{figure}[ht!]
\begin{center}
\includegraphics*[width=.23\linewidth]{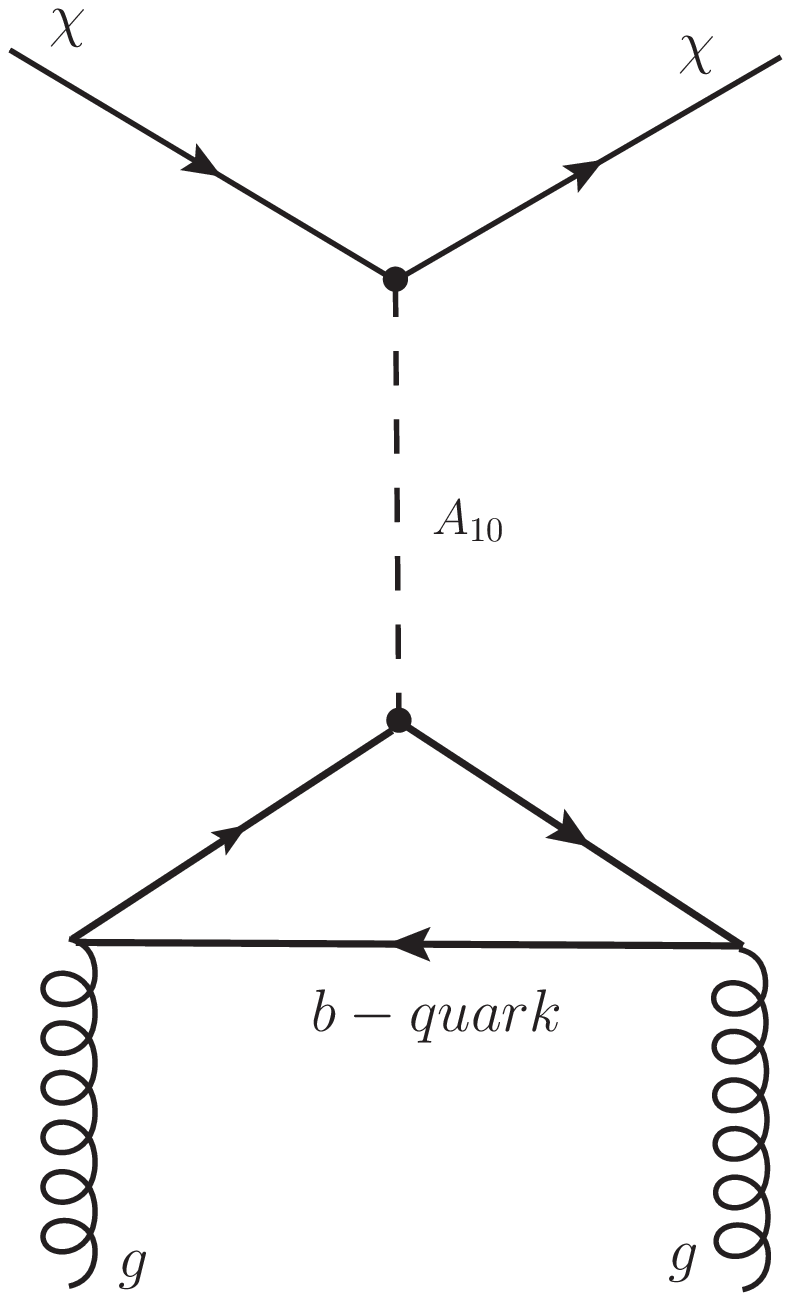}
\hskip 0.5cm
\includegraphics*[width=0.6\linewidth]{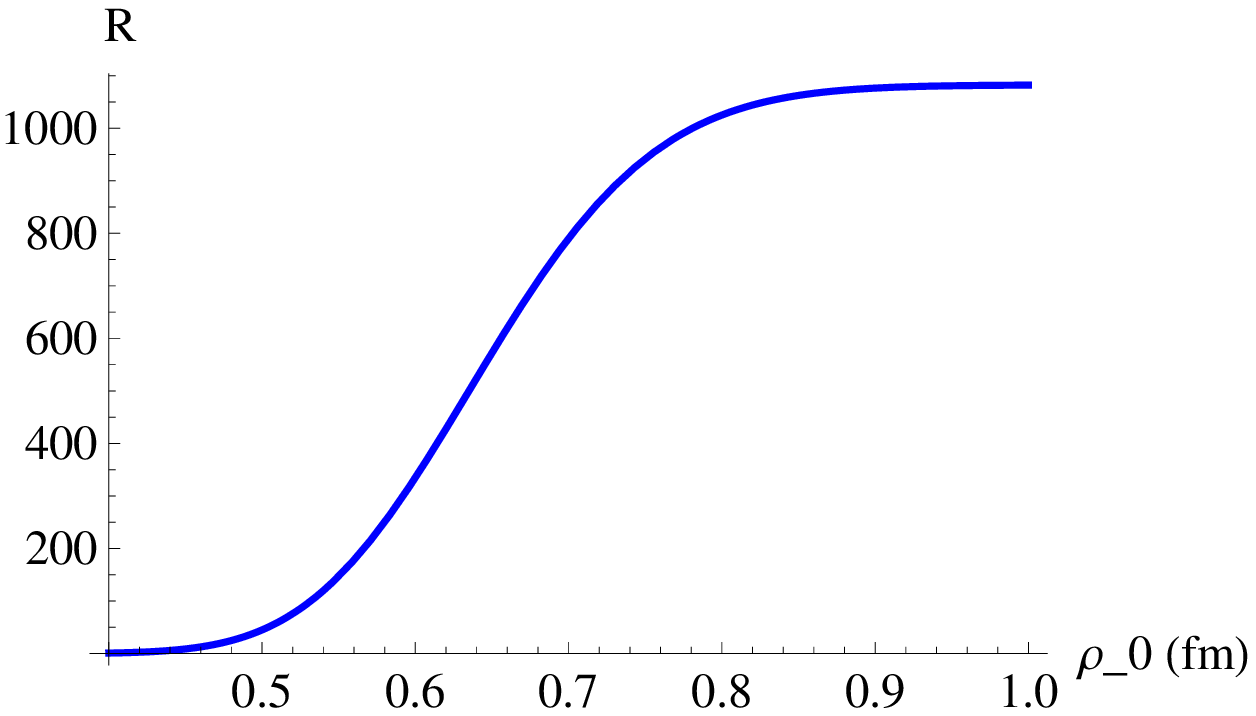}
\end{center}
\caption{Left: Contribution of a light CP-odd 
Higgs boson to the spin-dependent cross section
of WIMPs (e.g. a light neutralino $\chi$)
off nuclei via two gluons whose coupling
to nucleons would be enhanced by instanton-induced effects
as explained in the main text.  Right: The ratio 
$R$ defined in Eq.(4.3)
as a function of the upper limit
of the instanton size $\rho_0$ (in fm).
The curve is only indicative of a possible large enhancement
and numerical values from the curve should be taken with 
great care.} 
\label{fig:2}
\end{figure}

In fact,  direct detection through WIMP scattering off nuclei 
typically implies
momentum transfer $q$ of the order of
100 MeV, i.e. nuclear target recoil energies
of the order of tens of keV.

Nevertheless, momentum-dependent interactions have been put forward~
\cite{astro-ph/0408346,Chang:2009yt,Freytsis:2010ne} in order to
alleviate the tension between the DAMA signal and the null results
from other experiments. Specifically we focus on 
a scenario beyond the Standard Model, with
a light pseudoscalar Higgs boson $A_1$ acting
as a mediator, as shown in Fig.2 (left). 
A detailed discussion of the motivations for such
a scenario (in particular, a light neutralino $\chi$ as a
dark matter candidate, and a light CP-odd Higgs boson according to 
the NMSSM) can be found in \cite{Domingo:2008rr,Panotopoulos:2011si}
and references therein. 

At large $\tan{\beta}$, the bottom quark running in the triangle-loop
of the diagram of Fig.2 
should give the dominant contribution, leading to an effective
coupling of a (highly) virtual pseudoscalar $b\bar{b}$ state to 
gluons which, in turn, would couple to the target nucleon.

In view of our previous analysis of the $\eta_b \to p\bar{p}$ decay,  
we next examine the possibility
that the QCD vacuum might also play a non-trivial role in WIMP scattering
off nuclei, eventually causing a (huge) enhancement
of the spin-dependent cross section. The underlying reason
of such an enhancement comes from a larger number of instantons
involved in the gluon-quark interaction, 
whose size satisfies $\rho \lesssim 1/q \simeq 1$ fm. Hence
large values of $\rho_0$
correspond to low values of the momentum transfer to nucleons.

As already mentioned in section 2, instanton self-interactions 
should lead to the modification of
the (otherwise divergent) size distribution (2.1). 
We parametrize the new distribution as
\begin{equation}\label{eq:newdis}
n(\rho) \to n(\rho)\ e^{-c\rho^2}
\end{equation}
where $c=(b-4)/2\bar{\rho}^2$ \cite{Diakonov:1995qy}.
Setting numerical values, one finds $c \simeq 0.9$ GeV$^2$, also
in agreement with Eq.(4.20) of Ref.\cite{He:2009sb}. A similar
suppression factor, namely $\exp{(-2\pi\sigma\rho^2)}$, where
$\sigma = (0.44 \mathrm{GeV})^2$, was obtained in 
\cite{Shuryak:1999fe}.

In order to take into account all the $\rho$ powers
in the instanton-induced interaction from Eq.(\ref{eq:amplitude}),  
let us define 
\begin{equation}\label{eq:integral}
f(\rho_0)=\int_0^{\rho_0}\ d\rho\ \rho^{15}\ e^{-c\rho^2} 
\end{equation}
where $\rho_0$ denotes the instanton size upper limit set by
the typical energy scale of the process.

Notice that the instanton-induced $\chi$-nucleon cross section should
be proportional to $f(\rho_0)^2$, i.e. largely
depending on the instanton size range defined by $\rho_0$. 
In view of our previous assumption
on the enhancement of the $\eta_c \to p\bar{p}$
decay rate due to instanton effects, let us introduce the ratio 
\begin{equation}\label{eq:ratio}
R(\rho_0) = \biggl[\frac{f(\rho_0)}{f(\rho_c)}\biggr]^2
\end{equation}

In Fig.~2 (right) we plot $R(\rho_0$) as a function of
the upper limit $\rho_0$ of integration in Eq.(\ref{eq:integral}).
In spite of
many (and large) uncertainties (mainly due to the strong
dependence on a power of $\rho$) and the risky
exprapolation to low momentum transfer, we tentatively infer
from the curve that a very important enhancement 
of the cross section for spin-dependent 
$\chi$-nucleon cross section can occur.

\section{Summary}

Instantons, fluctuations of (non-abelian) gauge fields
representing topological changing tunneling transitions, yield
interactions between quarks and gluons 
which are absent in perturbation theory.
In this work we have first examined possible instanton-induced
effects in the decay of $\eta_b$ resonances into baryon pairs.
By rescaling the results experimentally found 
for $BF[\eta_c(1S)\to p\bar{p}]$
and  $BF[\eta_c(1S)\to \Lambda\bar{\Lambda}]$ we have been able
to make the order-of-magnitude estimate  
$BF[\eta_b(1S)\to p\bar{p}] \simeq BF[\eta_c(1S)\to p\bar{p}] 
\sim 10^{-7}-10^{-6}$, although with large uncertainties mainly
coming from the strong $\rho$ power dependence of the decay width
as can be seen from Eq.(\ref{eq:amplitude}).

On the other hand, we have 
examined possible instanton-induced
effects in WIPM scattering off nuclei, by extrapolation 
from the charmonium mass 
down to very small momentum transfer ($q \simeq \Lambda$)
implying large size instantons 
whose contribution is
cut off by an exponential factor. This possibility
covers different scenarios involving a light CP-odd
Higgs boson \cite{Panotopoulos:2011si}
or a light $Z'$ mediator (see e.g. \cite{Frandsen:2011cg}). 
We tentatively conclude that an important enhancement of
the spin-dependent cross section can be caused by
an instanton-induced interaction of gluons and nucleons. 
Phrased in other way, WIMP
scattering off nuclei would somewhat 
resemble (diffractive) proton-proton
collisions at high energy (see e.g. \cite{Kharzeev:2000ef}).  

Whether or not this kind of instanton-induced interaction associated to
helicity flip of the target nucleons is called to play a relevant
role in the description of dark matter scattering off nuclei, has 
to be investigated further. 

\subsection*{Acknowledgements}
The work by M.-A.S.-L. and R.P.-R. was partially 
supported by research grants FPA2011-23596
and GVPROMETEO2010-056. R.P.-R. also
acknowledges support from the 
academy researcher program of the Academy of
Finland, Project No. 130472. 
The work by N.K. was supported in part by South Korean Brain Pool Program 
of MEST,  by RFBR grant
10-02-00368-a and by Belarus-JINR grant.

\end{document}